# Detection of Advanced Malware by Machine Learning Techniques

Sanjay Sharma[1], C. Rama Krishna[2] and Sanjay K. Sahay[3]
[1]M.E. Scholar, Department of Computer Science and Engineering,
[2]Professor and Head, Department of Computer Science and Engineering,
[3]Assistant Professor, Department of Computer Science and Information System,
[1,2]National Institute of Technical Teachers Training and Research, Chandigarh, India
[3]BITS, Pilani, Goa Campus, India
[1]sanjay.cse@nitttrchd.ac.in, [2] rkc_97@yahoo.com, [3]ssahay@goa.bits-pilani.ac.in

**Abstract**

In today's digital world most of the anti-malware tools are signature based which is ineffective to detect advanced unknown malware viz. metamorphic malware. In this paper, we study the frequency of opcode occurrence to detect unknown malware by using machine learning technique. For the purpose, we have used kaggle Microsoft malware classification challenge dataset. The top 20 features obtained from fisher score, information gain, gain ratio, chi-square and symmetric uncertainty feature selection methods are compared. We also studied multiple classifiers available in WEKA GUI based machine learning tool and found that five of them (Random Forest, LMT, NBT, J48 Graft and REPTree) detect the malware with almost 100% accuracy.

## 1. Introduction

A program/code which is designed to penetrate the system without user authorization and takes inadmissible action is known as malicious software or malware [1]. Malware is a term used for Trojan Horse, spyware, adware, worm, virus, ransomware, *etc*. As the cloud computing is attracting the user day by day, the servers are storing enormous data of the users and thereby luring the malware developers. The threats and attacks have also increased with the increase in data at Cloud Servers. Figure 1 shows the top 10 windows malware reported by quick heal [2].

Malwares are classified into two categories - first generation malware and second generation malware. The category of malware depends on how it affects the system, functionality of the program and growing mechanism. The former deals with the concept that the structure of malware remains same, while the later states that the keeping the action as is, the structure of malware changes, after every iteration resulting in the generation of new structure [3]. This dynamic characteristic of the malware makes it harder to detect, and quarantine. The most important techniques for malware detection are signature based, heuristic based, normalization and machine learning. In past years, machine learning has been an admired approach for malware defenders.

In this paper, we investigate the machine learning technique for the classification of malware. In the next section, we discuss the associated work; section 3 describes our approach comprehensively, section 4 includes experimental outcomes and section 5 contains inference of the paper.

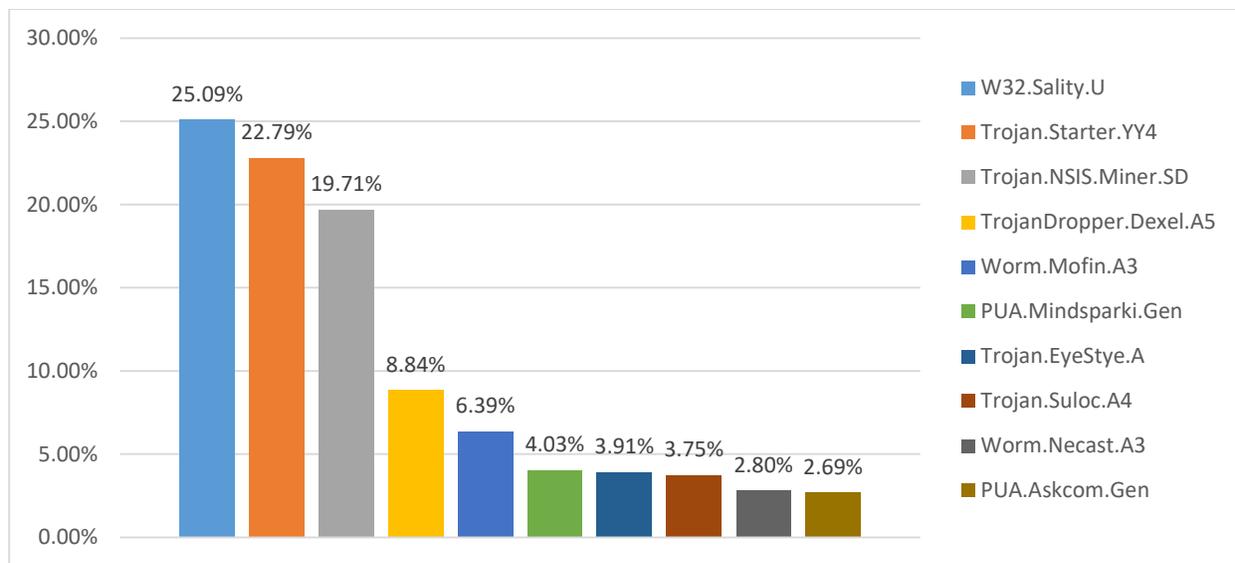

**Figure 1. Top 10 windows malware**

## 2. Related Work

In 2001 Schultz et al. [4] introduced machine learning for detection of unknown malware based on static features, for feature extraction author used PE (Program Executables), byte n-gram & Strings. In the year 2007, Danial Bilar [5] introduced opcode as a malware detector, to examine opcodes frequency distribution in malicious and non-malicious files. In year 2007, Elovici et al. [6] used Program Executable (PE) and Fisher Score (FS) method for feature selection and used Artificial Neural Network (5grams, top 300, FS), Bayesian Network (5-grams, top 300, FS), Decision Tree (5-grams, top 300, FS) , BN (using PE) and Decision Tree (using PE) and obtained 95.8 % accuracy. In the year 2008 Moskovitch et al. [7] used filters approach for feature selection. They used Gain Ratio(GR) and Fisher Score for feature selection and Artificial Neural Networks (ANN), Decision Tree (DT), Naïve Bayes (NB), Adaboost.M1 (Boosted DT and Boosted NB) and Support Vector Machine (SVM) classifiers and got 94.9 % accuracy.

In the year 2008 again, Moskovitch et al. [8] presented an approach in which they used n-gram (1,2,3,4,5,6 gram) of opcodes as features and used Document Frequency(DF), GR and FS feature selection method. They used ANN, DT, Boosted DT, NB and Boosted NB classification algorithms, out of this ANN, DT, BDT out-performed, preserving the low level of false positive rate.

In 2011 Santos et al. [9] inferred that supervised learning requires labelled data, so they proposed semi-supervised learning to detect unknown malware. In 2011, Santos et al. [10] again come with the frequency of the appearance of operational codes. They used information gain method for feature selection, and different classifiers, *i.e.* DT, k-nearest neighbor (KNN), Bayesian Network, Support Vector Machine (SVM), among them SVM outperforms with 92.92 % for one opcode sequence length and 95.90 for two opcode sequence length. In the year 2012, Shabtai et al. [11] used opcode n-gram pattern feature and to identify the best feature they used Document Frequency (DF), G-mean and Fisher Score method. In their approach, they used many classifiers, in which Random Forest outperforms with 95.146 % accuracy.

In 2016 Ashu et al. [12] presented a novel approach to identify unknown malware with high accuracy. They analyzed the occurrence of opcodes and by grouping the executables. Authors studied thirteen classifiers found in the WEKA machine learning tool out of them a Random forest, LMT, NBT, J48, and FT examined in depth and got more than 96.28% malware detection accuracy. In 2016 Sahay et al. [13] grouped executables on the base of malwares size by using Optimal k- means clustering algorithm, and these groups used as promising features for training (NBT, J48, LMT, FT and Random Forest) the classifiers to identify unknown malware. They found that detection of unknown malware by proposed the approach gives accuracy up to 99.11%.

Recently some authors worked on malware dataset released for kaggle dataset [14]. In the year 2016, Ahmadi et al. [15] took Microsoft malware dataset and used hex dump-based features (n-gram, Metadata, entropy, image representation and string length ) as well as features extracted from disassembled file (Metadata, Symbol frequency, opcodes, register, etc. ) and XGBoost classification algorithm. They reported ~99.8 % detection accuracy. In 2017 Drew et al. [16] used The Super Threaded Reference Free Alignment-Free Nsequence Decoder (STRAND) classifier to perform classification of polymorphic malware. In their approach, they presented ASM sequence model and obtained accuracy greater than 98.59 % using 10-fold cross-validation.

## 3. Methodology

To detect the unknown malware using machine learning technique, a flow chart of our approach is shown in fig. 2. It includes preprocessing of dataset, promising feature selection, training of classifier and detection of advanced malware.

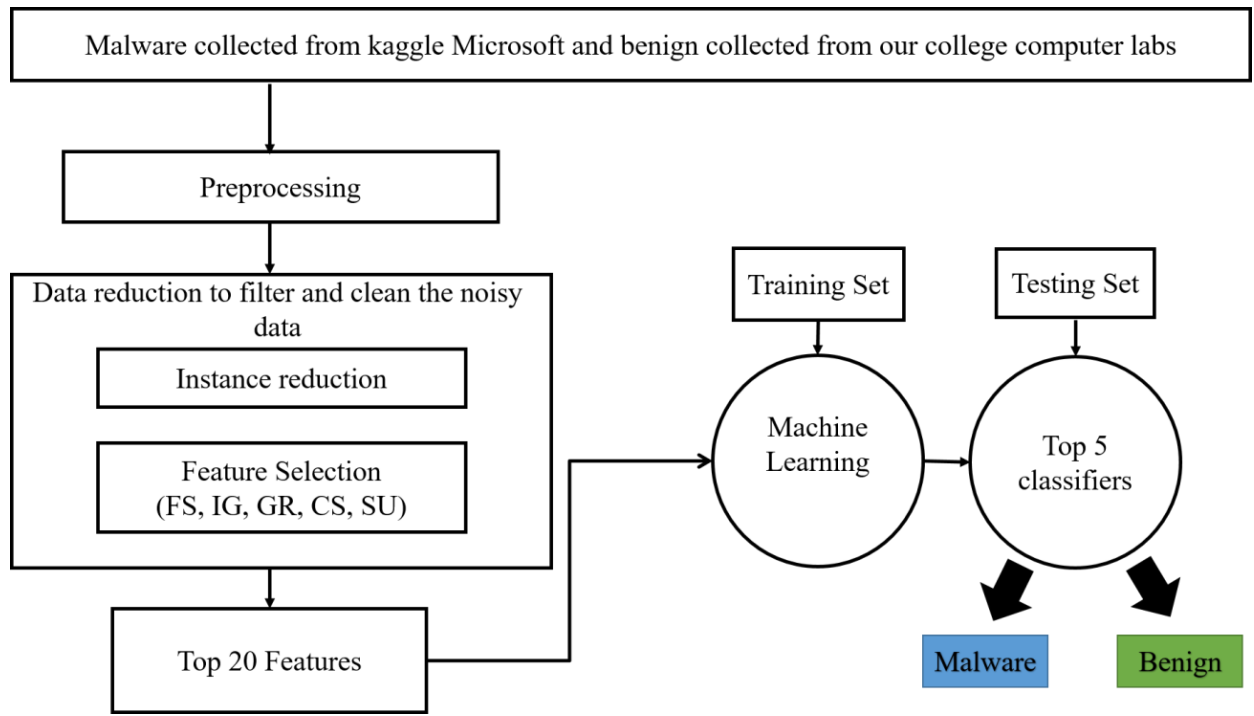

Figure 2. Flow Chart for Malware Detection

### 3.1 Building the dataset

Microsoft released approximately half terabyte for kaggle Microsoft Malware Classification Challenge (2015) [14] containing malware (21653 assembly codes). We downloaded malware dataset from kaggle Microsoft and collected benign programs (7212 files) for the windows platform (checked from virustotal.com) from our college's lab. In our experiment, we found that as dataset grows, there is an issue of scalability. This issue increases time complexity, storage requirement and decreases system performance. To overcome these issues, reduction of data set is necessary. Two approaches can be used for data reduction viz. *Instance Selection (IS)* and *Feature Selection (FS)*. In our approach, Instance Selection (IS) is used to reduce the number of instances (rows) in dataset by selecting most appropriate instances. On the other hand, Feature Selection is used for the selection of most relevant attributes (features) in dataset These two approaches are very effective in data reduction as they filter and clean, noisy data which results in less storage, time complexity and improve the accuracy of classifiers [17] [18].

## *3.2 Data Preparation*

From the earlier studies [12] we have found that opcodes contain a more meaningful representation of the code, so in proposed approach, we use opcodes as features. Malware dataset contains 21653 assembly codes of malware representation, a combination of 9 different families, i.e., Ramnit, Lollipop, Kelihos_ver3, Vundo, Simda, Tracur, Kelihos_ver1, Obfuscator.ACY, Gatak. Collected benign executables disassembled using *objdump* utility available in Linux system to get the opcodes.

In the malware dataset, we have found that maximum size of assembly code is 147.0 MB, so all the benign assembly above the 147.0 MB are not considered for the analysis. From earlier studies, we found that there are 1808 unique opcodes [12] so in our approach, there are 1808 features for machine learning. Then the frequency of each opcode in every malware and the benign file is calculated. After that in every malware and benign file total opcodes weight is calculated. Then it is noticed that there are 91.3 % malware file and 66 % benign file which contains opcodes weight below 40000. So to maintain the proportion of malware and benign all the files under 40000 weight is selected. After this step, 19771 and 4762 malware and benign files are left for analysis.

The next step is to remove noisy data from malware for that we have calculated the malware and benign files in the 500 intervals of opcodes weight. Those intervals in which there are no benign files, malware files are also deleted in that interval. In this way further intervals 100, 50, 10 and 2 of opcodes weights are created as shown in Fig. 3 to remove the noise from malware. Finally, dataset contains 6010 Malware and 4573 benign files.

| Opcode weight interval | no. of malwares | no. of benigns |
|---|---|---|
| 1-50 | 39 | 12 |
| 51-100 | 11 | 3 |
| 101-150 | 11 | 4 |
| 151-200 | 33 | 18 |
| 201-250 | 43 | 1 |
| 251-300 | 26 | 1 |
| 351-400 | 18 | 2 |
| 401-450 | 23 | 1 |
| 451-500 | 11 | 1 |
| 501-550 | 31 | 33 |
| 551-600 | 129 | 8 |
| 601-650 | 368 | 5 |
| 651-700 | 356 | 38 |
| 701-750 | 303 | 24 |
| 751-800 | 111 | 12 |
| 801-850 | 73 | 7 |
| 851-900 | 193 | 23 |

**Figure 3. Opcode Weight Interval Over Period of 50**

## 3.3 Feature Selection

Feature selection is an important part of machine learning. In proposed approach, there are 1808 features among them many do not donate to the accuracy and even decrease it. In our problem reduction of features is crucial to maintaining accuracy. Thus we first used *Fisher Score (FS)* [19] for feature selection and later four more feature selection techniques were also studied. The five feature selection method employed in this approach which functions according to the filters approach [20]. In this method, correlation of each feature with the class (Malware or benign) is quantified, and its contribution to classification is calculated. This method is independent of any classification algorithm unlike wrapper approach and allows to compare the performance of different classifiers. In this approach, *Fisher Score (FS), Information Gain (IG), Gain Ratio (GR), Chi- Square (CS) and Uncertainty Symmetric(US)* is used. Based on these feature selection measures we have selected top 20 features as shown in table 1.

**Table 1. Top 20 Features**

| Rank | Information Gain | Gain Ratio | Symmetrical uncertainty | Fisher Score | Chi Square |
|---|---|---|---|---|---|
| 1 | jne | jne | jne | je | jne |
| 2 | je | je | je | jne | je |
| 3 | dword | dword | dword | start | dword |
| 4 | retn | retn | retn | cmpl | retn |
| 5 | jnz | jnz | jnz | retn | jnz |
| 6 | jae | jae | jae | dword | jae |
| 7 | offset | offset | offset | test | offset |
| 8 | jz | movl | movl | cmpb | movl |
| 9 | movl | cmpl | jz | xor | jz |
| 10 | cmpl | jz | cmpl | jae | cmpl |
| 11 | int | movzwl | movzwl | movzwl | int |
| 12 | movzwl | movb | movb | ret | movzwl |
| 13 | movb | sete | sete | jbe | movb |
| 14 | sete | int3 | int3 | movl | sete |
| 15 | int3 | testb | testb | andl | int3 |
| 16 | testb | setne | cmpb | lea | testb |
| 17 | setne | cmpb | setne | cmp | cmpb |
| 18 | cmpb | andl | andl | testb | setne |
| 19 | andl | incl | incl | incl | andl |
| 20 | incl | movzbl | movzbl | setne | incl |

## 3.4 Training of the Classifiers

After the feature selection, next step is to find the best classifier for the detection of advanced malware. Next step is to compare different classifiers on FS, IG, GR, CS and US using top 20 features. We studied nine classifiers viz. Decision Stump, Logistic Model Tree (LMT), Random Forest, J48, REPTREE, Naïve Bayes Tree (NBT), J48 Graft, Random Tree, Simple CART available in WEKA. WEKA is an open source GUI based machine learning tool. We run all these classifiers on each feature selection technique using 10–fold cross-validation to train the classifiers. Fig. 4 shows the accuracy of each classifier concerning feature selection method. From the Fig. 4 it is clear that Fisher score method is best in among all and got accuracy 100 % in case of Random Forest, LMT, NBT and Random Tree. So in our proposed, Fisher Score performs better than other methods viz. Information Gain (IG), Gain Ratio (GR), Symmetrical Uncertainty and Chi-Square.

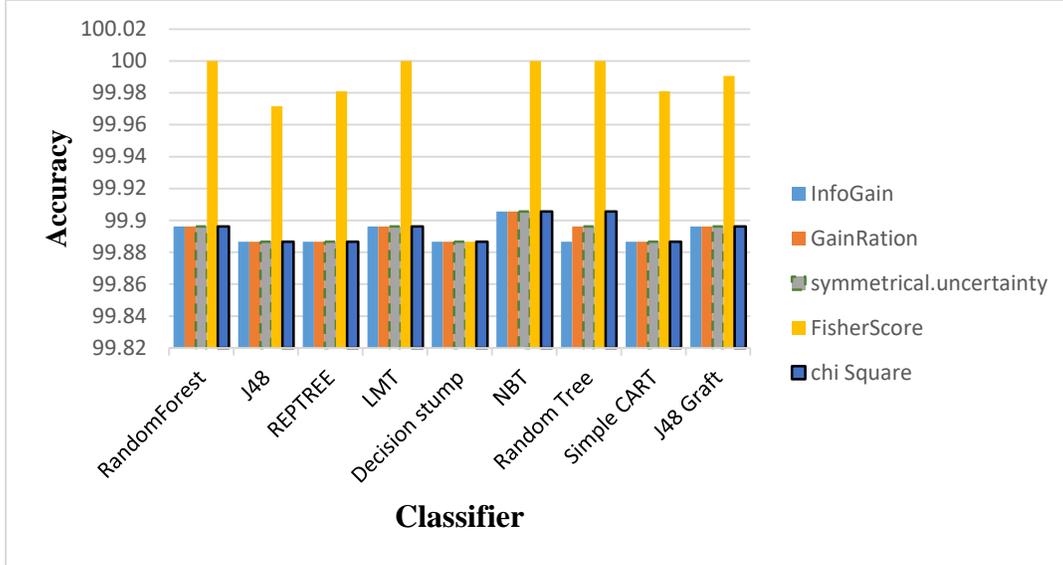

**Figure 4. Accuracy of Classifiers concerning different Feature Selection Methods**

## *3.5 Unknown Malware Detection*

In an earlier section, we have noticed that Random Forest, LMT, NBT, J48 Graft and Random Tree achieved maximum accuracy, so we selected these five classifiers for depth analysis. We have randomly selected 3005 malware and 2286 benign programs which are 50% of the overall dataset. Table 2 shows the results of top 5 classifiers.

## 4. Experimental Results

As mentioned in section 3, malware is already in assembly code only benign are disassembled.

Then opcodes occurrence is calculated for all malware and benign programs. In next noise from malware, data is removed by creating an interval of opcodes weight i.e. 500, 100, 50, 10, 5 and 2 for malware and benign files. Interval in which there are no benign files, malware files are deleted. To find the dominant features or to remove irrelevant feature we used five feature selection methods and found that there are 20 features which are dominating in the classification process. Fig. 4 shows that Fisher Score outperforms among five feature selection methods.

The analysis of top five classifiers viz. Random Forest, LMT, Random Tree, J48Graft and REPTree done in WEKA to find their effectiveness, regarding True Positive Ratio (TPR), True Negative Ratio (TNR), False Positive Ratio (FPR), False Negative Ratio and Accuracy, defined as

$$TPR = \frac{True\ Positive}{Total\ MAlware} \qquad TNR = \frac{True\ Negative}{Total\ Benign} \qquad FPR = \frac{False\ Positive}{Total\ Benign} \qquad FNR = \frac{False\ Negative}{Total\ Malware}$$

$$Accuracy = \frac{TP+TN}{TM+TB} \times 100$$

Where

True Positive: the no. of malware correctly detected.

True Negative: the no. of benign correctly detected.

False Positive: the no. of benign identified as malware.

False Negative: the no. of malware identified as benign.

Table 2 shows the result obtained by the top 5 classifiers. The study shows that the selected five classifiers accuracy is more or less same.

**Table 2. Performance of Top 5 Classifiers with Fisher Score Feature Selection Method**

| Classifiers | True Positive | False Negative | False Positive | True Negative | Accuracy |
|---|---|---|---|---|---|
| Random Forest | 100% | 0 | 0 | 100% | 100% |
| LMT | 100% | 0 | 0 | 100% | 100% |
| NBT | 100% | 0 | 0 | 100% | 100% |
| J48 Graft | 100% | 0 | 0 | 100% | 100% |
| REPTREE | 99.98% | 0.04% | 0.05% | 99.95% | 99.96% |

## 5. Conclusion

In this paper, we have presented an approach based on opcodes occurrence to improve malware detection accuracy of the unknown advanced malware. Code obfuscation technique is a challenge for signature based techniques used by advanced malware to evade anti-malware tools. Proposed approach uses Fisher Score method for the feature selection and five classifiers used to uncover the unknown malware. In proposed approach Random forest, LMT, J48 Graft, and NBT detect malware with 100% accuracy which is better than the accuracy (99.8%) reported by Ahmadi et al. (2016). In future, we will implement proposed approach on different datasets and will perform in the deep analysis for the classification of advanced malicious software.

**Acknowledgement**

Mr. Sanjay Sharma is thankful to Dr. Lini Methew, Associate Professor and Dr. Rithula Thakur Assistant Professor, Department of Electrical Engineering for providing computer lab assistance time to time.